# Global LCOEs of Decentralized Off-grid Renewable Energy Systems

Jann Michael Weinand[1,*], Maximilian Hoffmann[1], Jan Göpfert[1], Tom Terlouw[3], Julian Schönau[1], Patrick Kuckertz[1], Russell McKenna[2,3], Leander Kotzur[1], Jochen Linßen[1], Detlef Stolten[1,4]

[1]Forschungszentrum Jülich GmbH, Institute of Energy and Climate Research – Techno-economic Systems Analysis (IEK-3), 52425 Jülich, Germany

[2]Chair of Energy Systems Analysis, Institute of Energy and Process Engineering, ETH Zurich, Switzerland

[3]Laboratory for Energy Systems Analysis, Paul Scherrer Institute, Switzerland

[4]RWTH Aachen University, Chair for Fuel Cells, Faculty of Mechanical Engineering, 52062 Aachen, Germany

*Corresponding author: Jann Michael Weinand, j.weinand@fz-juelich.de, +49 175 4985402

**Abstract**

Recent global events emphasize the importance of a reliable energy supply. One way to increase energy supply security is through decentralized off-grid renewable energy systems, for which a growing number of case studies are researched. This review gives a global overview of the levelized cost of electricity (LCOE) for these autonomous energy systems, which range from 0.03 $_{2021}$/kWh to over 1.00 $_{2021}$/kWh worldwide. The average LCOEs for 100% renewable energy systems have decreased by 9% annually between 2016 and 2021 from 0.54 $_{2021}$/kWh to 0.29 $_{2021}$/kWh, presumably due to cost reductions in renewable energy and storage technologies. Furthermore, we identify and discuss seven key reasons why LCOEs are frequently overestimated or underestimated in literature, and how this can be prevented in the future. Our overview can be employed to verify findings on off-grid systems, to assess where these systems might be deployed and how costs evolve.

**Highlights**

- Global overview of 161 studies on LCOEs of decentralized off-grid energy systems since 1990
- Studies using simulation models tend to oversize system components
- LCOEs for 100% renewable energy systems have decreased by 9% annually between 2016 and 2021
- Most case studies were conducted for India (22%), Iran (7%), China (7%), Nigeria (5%) and Canada (4%)

**Word count (without table and figures):** 3824

# 1 Introduction

Recent events have reduced the otherwise steadily increasing annual percentage of the global population with access to electricity for the first time in years [1]. Due to long distances to grid infrastructure, off-grid renewable energy systems are economically viable options to provide larger electricity access in developing regions like sub-Saharan Africa [2–4]. Even in industrialized countries with nationwide electrification, many local communities are striving for autonomous energy systems with 100% renewable energies [5–7], often motivated by economic, environmental and/or social reasons [8]. Decreasing costs for renewable energy technologies [9,10] as well as current uncertainties in the supply with fossil fuels and discontinued supply [11] will probably further increase these incentives for energy autonomy.

For several years the feasibility of 100% renewable energy systems has been controversially discussed [12–14] and there have been some insights into how these systems could be implemented [15–17]. Other relevant studies include recent bibliometric analyses of 100% renewable energy systems [18], comprehensive reviews of the history and future of 100% renewable energy systems [19], reviews of 100% renewable energy scenarios on islands [20], and reviews of best practices and potential improvements for modeling such energy systems [2]. While the majority of these studies focusses on national energy systems, the latter two studies partly address the levelized cost of electricity (LCOEs) for decentralized energy systems. In Meschede et al. [20] this is only dealt with sporadically, whereas Weinand et al. [2] analyze the LCOEs for decentralized autonomous energy systems in a more detailed way. However, since the publication of the latter study, the number of studies on decentralized energy autonomy has increased considerably (see Section 2) and the costs are not discussed in detail because the study focuses more on modeling aspects. Therefore, with the present review, we intend to provide an up-to-date overview of the cost (developments) for decentralized autonomous energy systems and give reasons for overestimation and underestimation of LCOEs.

# 2 Methods

We have identified 730 studies between 1990 and 2021 in the literature database Scopus[1] [21]. Through a manual check of titles, abstracts and full texts, we identified 228 articles which address decentralized energy autonomy in small regions such as villages, municipalities, islands, or cities. This number of articles nearly doubled between 2020 and 2021 with 105 new studies in this period. 161 of the 228 articles [22–182] specify LCOEs for autonomous energy systems (see Figure 1), of which 83 studies were published until 2019 and were previously identified by Weinand et al. [2]. Energy system analyses for individual residential, commercial, or industrial buildings/applications as well as analyses of large regions such as federal states, entire countries, or continents were excluded here. All economic cost values stated below are inflation adjusted and refer to the year 2021. Furthermore, studies with LCOEs above 1 $_{2021}$/kWh are excluded in the following analysis (see explanations in Section 3.1).

---

[1] Search query taken from Weinand et al. [2]: TITLE-ABS-KEY ("energy system" AND ("simulation" OR "modelling" OR "optimisation" OR "analysis") AND ("region" OR "municipalities" OR "municipality" OR "communities" OR "community" OR ("district" AND NOT "district heating") OR "city" OR "cities" OR "town" OR "remote") AND ("off-grid" OR "off grid" OR ("100%" AND "RE") OR ("100%" AND "renewable") OR "100%-renewable" OR ("energy" AND "autonomy") OR ("energy" AND "autarky") OR ("energy" AND "self-sufficiency") OR ("energy" AND "self-sufficient") OR "energy independent" OR "stand-alone" OR "energy autonomous" OR "island system")) AND (LIMIT-TO (DOCTYPE,"ar")) AND (LIMIT-TO (LANGUAGE , "English"))

# 3  Results

The inflation-adjusted LCOEs calculated by the 161 case studies range from 0.03 $\$_{2021}$/kWh in Alotaibi & Eltarnaly [36] (Saudi-Arabia) to 0.99 $\$_{2021}$/kWh in Rehman et al. [161] (Pakistan), with a total mean value of about 0.35 $\$_{2021}$/kWh (median is 0.29 $\$_{2021}$ and mode is 0.24 $\$_{2021}$, see Figure 1). Since 2016, the mean LCOEs for autonomous energy systems have decreased from 0.33 $\$_{2021}$/kWh (<100% renewable, i.e., including fossil fuels) and 0.54 $\$_{2021}$/kWh (100% renewable) on average by 4% and 9% per year to 0.23 $\$_{2021}$/kWh and 0.29 $\$_{2021}$/kWh in 2021, respectively. In all articles that consider both hybrid renewable-fossil-fuel systems and 100% renewable systems, the latter are on average 24% more costly. However, all hybrid systems include large shares of renewables and due to the above-mentioned stronger cost degression for 100% renewable systems, the cost deviation could progressively diminish.

Most studies in the research field of energy system analysis originate from the USA, China, United Kingdom, Germany and Italy [183], however, most of these countries are underrepresented in the 161 case studies on off-grid systems. Among the case studies that explicitly mention LCOEs, most were conducted for India (22%), Iran (7%), China (7%), Nigeria (5%) and Canada (4%). While 3% of the studies were conducted for German and 1% for Italian regions, no case studies were published for the USA or the United Kingdom. In some countries such as Spain [89], Germany [134] and New Zealand [137] with comparatively high electricity prices (cf. Figure 1), the calculated LCOEs for off-grid systems are partly below the household electricity prices (which also contain taxes and levies) in December 2021 of 0.32 $\$_{2021}$/kWh, 0.34 $\$_{2021}$/kWh and 0.19 $\$_{2021}$/kWh, respectively [184].

Of the 161 case studies, 100 consider 100% renewable energy systems without fossil fuels. The majority of these studies (63%) applied the HOMER (Hybrid Optimization of Multiple Energy Resources) or HOMER Pro simulation models. The HOMER model is a widely used open-source software tool for designing microgrid systems. Developed by the National Renewable Energy Laboratory (NREL), it is used to evaluate the technical and economic feasibility of integrating different energy sources, such as solar, wind, and energy storage, into a microgrid. The model considers inputs such as weather, load profiles, and equipment performance to determine the optimal configuration of a microgrid system. Other studies used the optimization models RE³ASON [177,178], Offgridders [58], LINGO [101], ISLA [143] and IREOM [102], the simulation models H$_2$RES [112], and EnergyPLAN [72] or metaheuristics like particle swarm optimization [95,115,151,159], genetic algorithms [96,115,140,154] or discrete harmony search [62]. Furthermore, Kumar & Saini [115] compare nine different metaheuristics for the energy system optimization of five un-electrified villages in India and demonstrate that the Salp Swarm Algorithm converges most efficiently.

In the next sections, we analyze why some studies overestimate (Section 3.1) or underestimate (Section 3.2) the costs of 100% renewable off-grid energy systems and how this could be improved in the future. Thereby, the focus lies on the 100 case studies with 100% renewable energy systems.

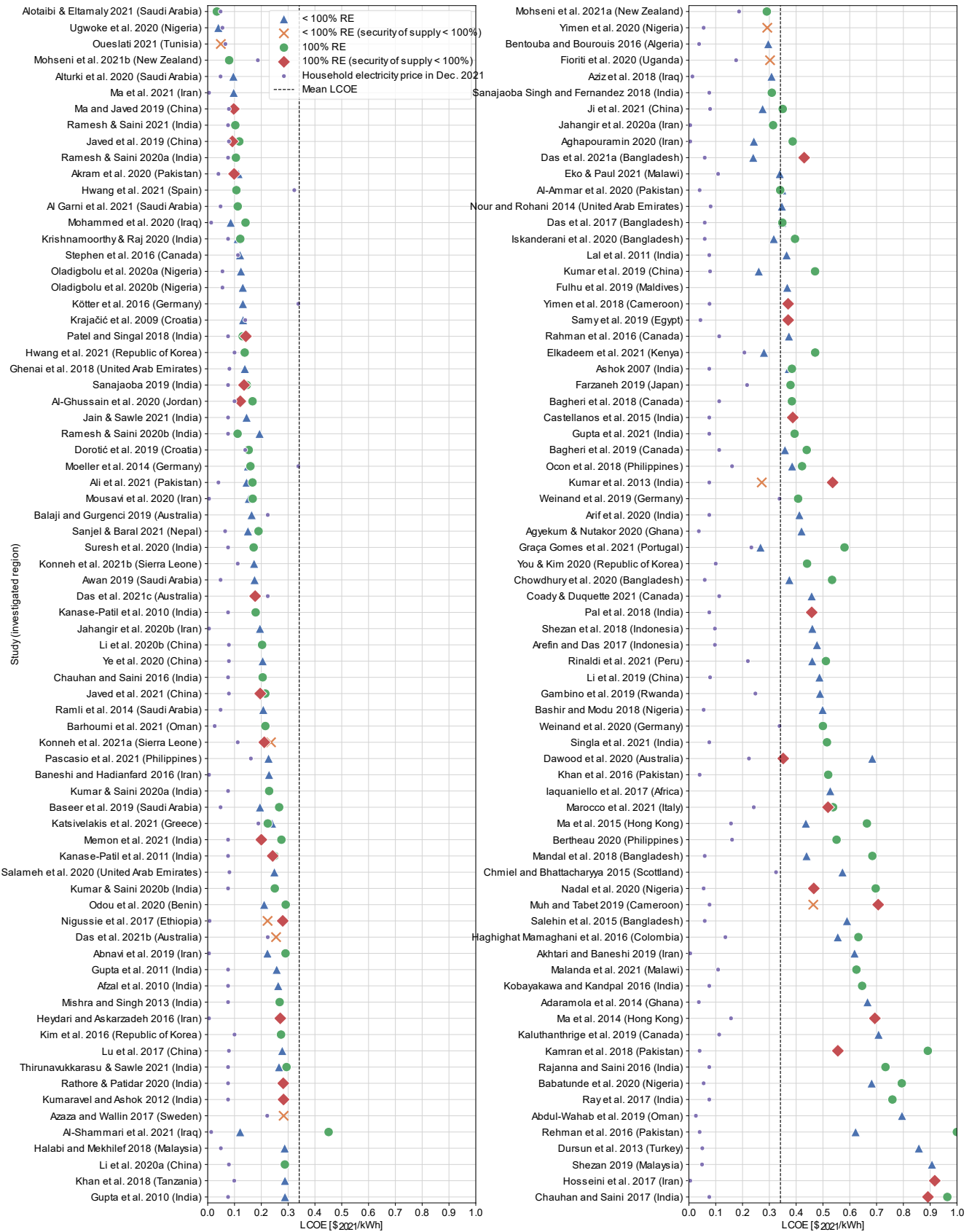

*Figure 1: Inflation-adjusted levelized cost of electricity (LCOE) for case studies on off-grid energy systems. The studies are sorted by mean LCOEs of all considered systems. Some hybrid systems consider fossil fuels and renewables (<100% RE) and some case studies incorporate only 100% renewable based systems (100% RE). The open-source Currency Conversion for Python (CuCoPy) [185] package was developed for this research and provides methods for exchanging currencies and adjusting monetary values for inflation. Its scope of application ranges from 1960 to 2021. Exchanging a value between currencies is done by dividing the target currency's exchange rate by the initial currency's exchange rate and multiplying the resulting quotient by the initial value. Likewise, adjusting for inflation is done by dividing the*

country's consumer price index (CPI) at the starting date by its CPI from the target date. Most exchange rates and consumer price indices were provided by the World Bank Group and used under the CC BY 4.0 license [186]. The exchange rate used for converting Indian Rupees to U.S. Dollars in 2021 was not included in the data provided by the World Bank Group and was instead calculated by averaging the monthly exchange rate of Indian Rupee against U.S. Dollar provided on pages 104 and 105 in the "Economic Survey 2021-2022 Statistical Appendix" conducted by the Reserve Bank of India and published by Union Budget (India) [187]. In a few studies [40,98,147], LCOEs were given, but it was not clear for which country the case studies were conducted. Since it is not possible to adjust for inflation and no household electricity price can be stated for comparison, these studies are not included in the figure. The household electricity prices all electricity bill items, such as the distribution and procurement costs, a variety of environmental and fuel costs, and taxes [184]. The right diagram is a continuation of the left diagram.

## 3.1 Reasons for overestimating LCOEs

While off-grid systems are generally associated with higher costs to meet load at all times of the year, a few studies show very high LCOEs, some above 1 $\$_{2021}$/kWh. Some LCOEs are particularly high due to high inflation in the countries studied, e.g., in the study from Askari & Ameri [44] for Iran from 2009. In the following, we provide an overview of some of the key drivers of why LCOEs have been overestimated in some studies.

**Investment decisions as model input.** Some studies using the energy system model HOMER present poor dimensioning of the autonomous energy system components. Chauhan et al. [63], for example, install an oversized hydro power plant in each of their scenarios and the 100% renewable energy system results in 98% excess electricity per year and LCOEs of 2.99 $\$_{2021}$/kWh. Similarly, in Bashir & Modu [55], Rahman et al. [153] and Chang et al. [60], the energy systems also show 65%-92% excess electricity due to large oversizing of system components. The problem with these studies is that due to the high combinatorial complexity of combined investment and dispatch optimization models [188], simulation models like HOMER are applied instead. This means that the dimensioning of the system components has to be done in advance and is not optimized within the model, which requires in-depth knowledge of the analyzed systems. While it is possible to achieve comparable results with simulation approaches [189], an application of advanced models for investment and dispatch optimization should be carefully considered in the future to avoid overestimation of costs.

**Ignoring technology cost degressions.** Many articles did not adjust their cost assumptions to real developments. Especially in the last years, the mean of the assumed costs for PV, onshore wind and battery storage in the studies is significantly above global cost trends (see Figure 2). Some notable examples include high PV costs of 2,500 $\$_{2021}$/kW in the article by Malanda et al. [129] from 2021, 4,200 $\$_{2021}$/kW in You & Kim [182] from 2020 or 5,800 $\$_{2021}$/kW in Baseer et al. [54] from 2019. Particularly high wind or battery costs are found in Malanda et al. [129] from 2021 with 6,000 $\$_{2021}$/kW for onshore wind or in Chang et al. [60] from 2021 with 1,700 $\$_{2021}$/kWh for battery storage. The peaks in 2017 for wind turbine and battery costs are related to the fact that only two studies report costs and these are relatively high: the high maximum costs for batteries and wind turbines based on Hosseini et al. [88] are related to the strong inflation in Iran, and the high minimum cost for onshore wind are related to the cost assumption of about 5,100 $\$_{2021}$/kW in Das et al. [69]. Since inflation-adjusted technology costs are compared with global cost developments in Figure 2, these do not necessarily coincide. Still, this reveals that the researchers' estimates of costs tend to be pessimistic. Cost developments and influences could, for example, be covered by sensitivity analyses, but generally only few to no studies conduct these analyses with regard to techno-economic parameters. An exception is Nadal et al. [140] which comprehensively investigates ranges of capital and operational expenditures, replacement times etc. for PV, electrolyzers and batteries. They show for a microgrid in Nigeria that capital costs of PV and capacity loss of batteries are among the most influential parameters on LCOEs, which again illustrates the importance of sound cost choices.

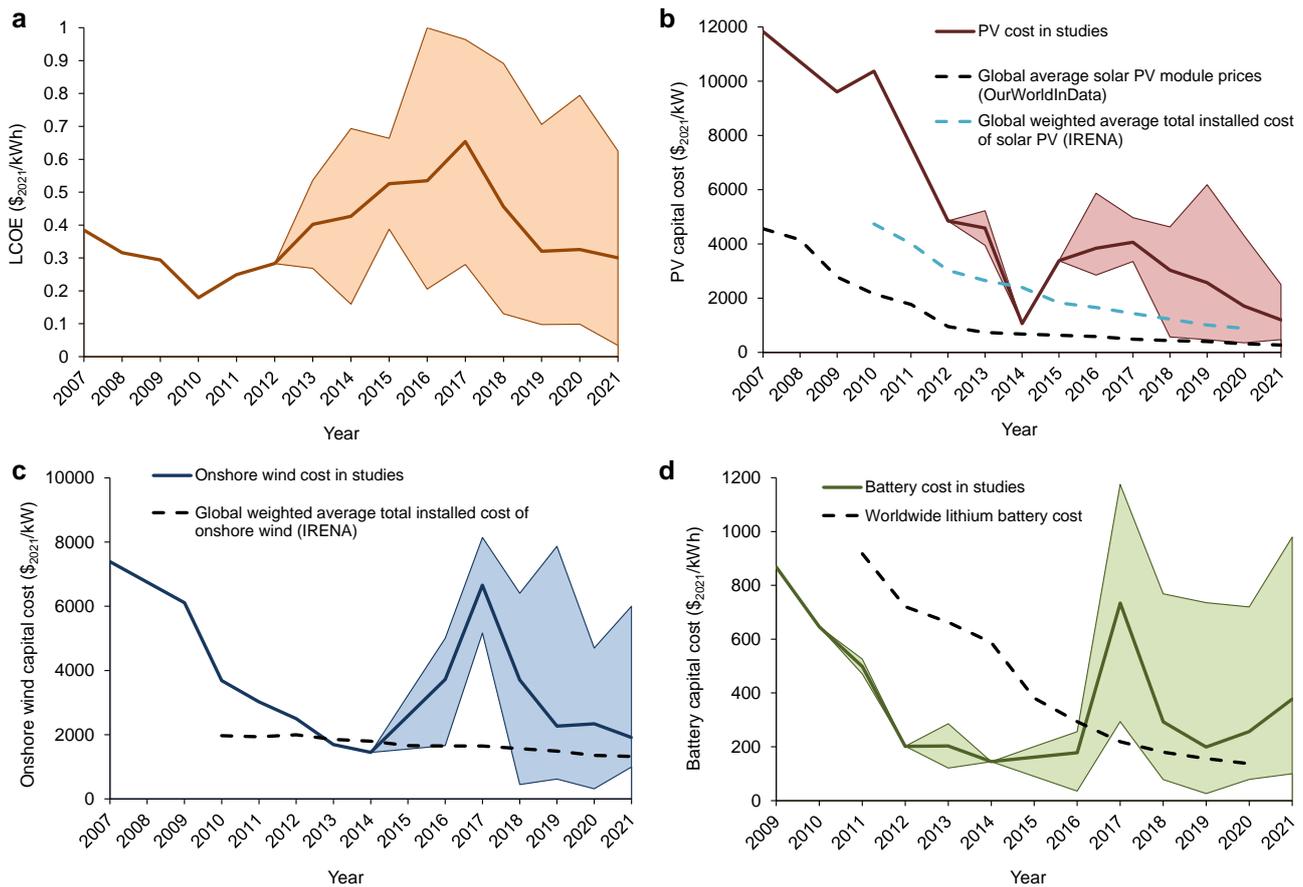

*Figure 2: Inflation-adjusted LCOEs (a) in 100 of the 161 studies, which consider 100% renewable energy systems without fossil fuels. PV capital cost (b), onshore wind capital cost (c) and battery capital cost (d) are only indicated in 89 of the 100 studies, i.e., 11 studies do not give information on costs. Due to their large impact on the cost curves, some very large outliers have been removed from a, b and c, see main text. The curves indicate the mean values among all studies and the area around them show the range between upper and lower extreme values. If no area surrounds the curve of mean values in a specific year, this means that either only one study was published in this year, or all used the same cost value. In panels a, b and c, the inflation-adjusted costs from the studies are also compared to real costs based on global averages. The global costs for PV are based on Refs. [190,191], for wind on Ref. [191] and for batteries on Ref. [192].*

**Neglecting technology options.** Not all articles consider comprehensive technology options. While solar PV is considered in all 100 studies and batteries in almost all articles (92%), this is not the case for other technologies (see Table 1). The importance of considering technologies comprehensively is shown by the fact that including onshore wind, hydro power, batteries or hydrogen storage, and fuel cells plus electrolyzers could reduce LCOEs on average between 15 to 36% (Table 1). This finding is in line with other research: a recent article has shown that neglecting onshore wind in municipal renewable energy systems leads up to about 0.08 $_{2021}$/kWh higher LCOEs for energy systems by 2050 [193]. Other studies show that incorporating base load capable technologies such as deep geothermal energy could also significantly reduce the cost of decentralized energy systems [177,178,194]. Particularly in off-grid energy systems, unconventional but potentially beneficial technologies and measures should also be incorporated in the future, e.g., higher shares of district heating [195,196] or the integration of large-scale hydrogen production [197].

Table 1: Impact of neglecting specific technologies on LCOEs for 100% renewable off grid energy systems in 25 case studies published in 2021.

|  | Wind power | Hydro power | Batteries | Elecrolyzers, fuel cells and hydrogen storage |
|---|---|---|---|---|
| Share of studies not including this technology [%] | 20 | 88 | 4 | 80 |
| Mean LCOE increase if not included [%] | 36 | 24 | 30 | 15 |

## 3.2 Reasons for underestimating LCOEs

There are also some key drivers, which could have led to an underestimation of LCOEs and will be discussed below.

**Neglecting grid integration.** Costs for integrating variable renewables into energy systems are small at low penetration of renewables, but can rise sharply at high penetrations [198,199]. Parts of the system LCOEs for integrating renewables are *profiling costs* for dispatchable generation to meet the residual demand, *balancing costs* to balance forecast and actual non-dispatchable generation, and *grid costs* for grid reinforcements and extensions to integrate the renewable generators in the network [200,201]. While in the case studies on 100% renewable energy systems the balancing costs are included and the profiling costs are at least partially included, the grid costs are neglected with very few exceptions (see Moeller et al. [134] or Weinand et al. [177,178]). Many recent articles show that LCOEs are underestimated if not all system LCOE aspects are considered: for example, Chen et al. [202] show for China that the traditional LCOE approach underestimates wind generation costs by about 15% compared to a system cost approach. Furthermore, McKenna et al. [203] demonstrate in an onshore wind potential analysis for Great Britain that taking grid connection costs into account doubles the cost of a wind farm on average. Veronese et al. [204] derive similar conclusions for solar PV in the future Italian energy system revealing that the system LCOEs are on average 50% higher than in usual LCOE analyses. Thus, future studies on 100% renewable energy systems should attempt to incorporate all components of system LCOEs.

**Applying hourly resolution.** Das et al. [67] use the HOMER model to show for a PV/Wind energy system with lithium-ion batteries in a remote community in Australia, that the temporal resolution of the model has a negligible effect on the LCOEs. Their results show that the LCOEs decrease with lower temporal resolution from about 0.33 $\$_{2021}$/kWh at a minute resolution to 0.32 $\$_{2021}$/kWh at an hourly resolution. For that reason, Das et al. decide for an hourly resolution given a smaller computational load. To the best of the authors' knowledge, the remaining works subject to this review focus on hourly resolution exclusively. Potential reasons are the generally better availability of hourly resolved data bases on the one and the moderate required model runtimes on the other hand, but also software-related restrictions as more than 50% of the reviewed publications rely on the software HOMER or HOMER Pro. These models use a hybrid approach of optimization and simulation to design near-optimal, but reliable systems, which may distort the impact of different temporal resolutions.

Purely optimization-based capacity expansion models are well-known to underestimate real system costs at coarser temporal resolutions [205] due to unintentional peak-shaving of the duration curves resulting from averaging [206] and to thereby undersize system capacities, which leads to operationally infeasible system designs [207]. This effect is particularly strong for small and isolated renewable energy systems, which cannot use grid connections or the superposition of multiple demand profiles to level out demand peaks or supply troughs, leading to significantly higher overcapacities if the temporal resolution is increased [208]. Furthermore, the cost increase is degressive with higher temporal resolutions and therefore it is highly model-dependent whether the impact of an increased temporal resolution can be neglected or not. For that reason, different optimization-based publications focusing on different model scopes have arrived at different conclusions with respect to the impact of sub-hourly model resolutions: for the cost-optimal design of a hybrid municipal energy system with 250 households comprising photovoltaics (PV) and combined heat and power (CHP), Kools et al. [209] conclude that higher temporal resolutions lead to slightly higher load losses (3% for minutely resolution, 2% for hourly resolution) and smaller PV capacities, but that the impact is small and should therefore be omitted for the sake of computational tractability. Harb et al. [210] arrive at a similar finding that the overall cost underestimation of less than 1% in hourly energy system optimizations of a small neighborhood compared to quarter hourly resolution is negligible, but the general trend that higher resolutions lead to higher costs and smaller cost-optimal shares of non-dispatchable renewables (if dispatchable fossil sources are available) holds true as well.

Overall, the impact of sub-hourly resolved time steps on overall system costs likely remain small or moderate, but that the systematic assessment of this aspect is too widely neglected to derive general conclusions. Especially with respect to the relative frequency of outage or lost load with usually very small percentage values, the impact may be considerable for 100% renewable off-grid systems.

**Risk of social opposition.** The vast majority of articles contain pure techno-economic analyses. Only a few studies combine this with multi-attribute [109] or multi-criteria [75,114] decision making to include preferences of stakeholders in the evaluation of energy systems. The disregard of social acceptance could lead to technically and economically optimal energy systems from a theoretical perspective, which cannot be implemented in reality, as decision-makers might reject certain technologies. Especially for onshore wind, the opposition of local inhabitants towards turbines due to landscape impacts [203,211,212] or disamenities [213,214] may be particularly strong and lead to higher system costs. Since many aspects regarding the techno-economic feasibility of off-grid renewable energy systems have already been extensively studied in the past, future studies should increase their efforts to incorporate more non-technical aspects in energy system analyses [215,216].

**Transformation versus overnight expansion.** With only four exceptions that consider off-grid energy systems in a multi-year transformation [72,112,177,178], nearly all studies (96%) consider so-called overnight pathways, i.e., only the cost-optimal final state is planned for the energy system, but not the path leading there. Especially expansion rates of renewable energies as well as retrofit rates of buildings can have a major impact on costs and $CO_2$ emissions in decentralized energy systems [194] and could be limited by available material and craftsmen. Using a multi-year transformation planning together with model-endogenous technology learning could also avoid stranded investments due to installing technologies that are not needed in the future energy system [217].

## 3.3 Considerable impact of discount rate on LCOE

Another significant influence on the LCOE in energy system analyses can arise from the choice of the discount rate, which is very country specific [218] and ranges from 0.3% for a case study in Japan to 18% for a case study in Iran (Figure 3). Some studies also examine the effect of discount rate on LCOE for regional energy system case studies in Bangladesh [130], Canada [50], Cameroon [139], India [155], and China [123]. Thereby, the studies show that a 10% increase in the discount rate increases the LCOE by about 3-6%. Due to the higher specific investment costs and lower operating costs of renewables, the discount rate has a particularly high impact in renewable energy systems. For a hybrid off-grid energy system with a renewable penetration of only about 20%, Rahman et al. [153] demonstrate that an increase in the discount rate of 20% has a negligible impact on costs (+ 0.1%). In future studies on 100% renewable energy systems, the choice of the discount rate should be made very carefully to avoid underestimation or overestimation of system costs.

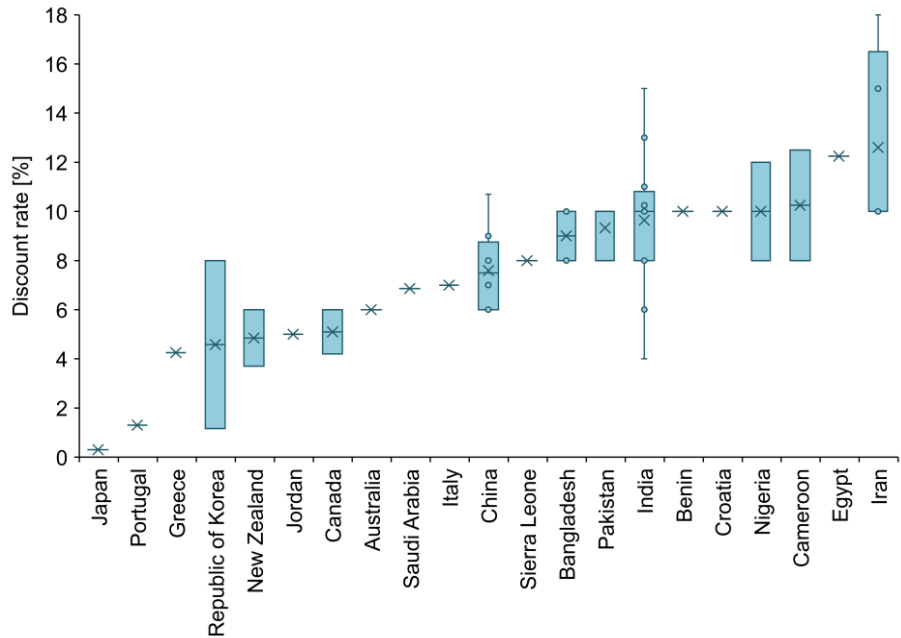

*Figure 3: Box plots of discount rates in the 59 articles on 100% renewable off-grid energy systems, classified by country in which the case study was investigated. 41 articles on 100% renewable off-grid energy systems do not state the value of the discount rate.*

# 4    Discussion

This review reveals the decrease of costs for decentralized off-grid renewable energy systems due to technological progress and cost degression. Recent global energy, health and geopolitical crises and the associated rise in retail energy prices could make off-grid energy systems worthwhile even in certain regions of industrialized countries. As has been shown for a few countries, the household electricity price is already higher than the LCOEs calculated in some case studies for off-grid energy systems.

Additionally, we have identified seven key reasons that lead to a systematic overestimation or underestimation of costs in the model calculations. To avoid an overestimation of LCOEs, future studies should carefully size energy technologies in simulation models (1), integrate all recent cost developments (2) and include potentially beneficial technology options from the analysis (3). To prevent underestimation of costs, integration costs should be accounted for (4), higher temporal resolutions should be applied in combination with time series aggregation approaches (5), social opposition to certain technologies in the regions studied should be addressed (6), and pathways for the transformation of energy systems should be planned instead of only planning the final state of the systems (7).

Further suggestions have recently been developed by a group of experts. As an energy system reaches 100% renewable energy, the necessary balance between supply and demand usually leads to a highly nonlinear increase in costs, mainly due to seasonal mismatches [17]. Since reaching the last 10% to achieve a completely renewable energy supply is especially challenging, the group of experts introduced six strategies for this [16]: building more variable renewable energy together with transmission and diurnal storages (1), installing other base-load capable renewable energy technologies like geothermal energy, hydropower or biopower (2), deploying nuclear plants as well as fossil-based ones with carbon capture (3), using seasonal storage by hydrogen, storage and re-electrification (4), employing carbon dioxide removal like bioenergy with carbon capture and storage (BECCS) or direct air carbon capture and storage (DACCS) (5) or intensifying demand-side measures like demand response or demand flexibility (6). While some of these strategies are more suited for large centralized energy systems (e.g., installing nuclear power plants) or fully decarbonized energy systems (BECCS and DACCS), they are consistent with our recommendation to exploit all available technological options to achieve 100% renewable energy systems in the future in a cost-effective way.

While we sought to present the LCOEs of off-grid regions in various countries as comparable as possible by adjusting for inflation, the heterogeneity of regions [219] means that the LCOEs between studies can never be completely comparable. In addition, in the discussion on system LCOEs in Section 3, we indicated that LCOEs may not be the best and most comprehensive metric to compare energy systems. Recently, researchers have introduced a new metric called the Cost of Valued Energy (COVE) to better evaluate energy systems with high shares of renewable energy. The COVE relies on system costs in relation to spot market revenue on an annual basis and thus takes into account not only the economic impact of supply vs. demand but also of cost vs. revenue [220]. Although spot markets could be irrelevant in decentralized off-grid energy systems, this highlights once again the need for novel metrics to compare energy systems.

Besides cost considerations, off-grid energy systems should be assessed by means of environmental metrics and social aspects to achieve a more thorough energy systems analysis. Life cycle assessment can be used to quantify the environmental impacts of a product, service, or energy system over the entire life cycle (including the manufacturing, operation, and end-of-life phase), considering environmental impacts beyond greenhouse gas (GHG) emissions [221]. Thus, life cycle assessments can identify potential trade-offs between costs, greenhouse gas emissions, and other environmental burdens [197,221]. Compared to incumbent fossil fuel-based systems, off-grid energy systems usually exhibit lower life cycle greenhouse gas emissions due to the integration of low-carbon energy sources, such as solar PV and wind. However, the manufacturing of off-grid energy systems can result in environmental burden shifting, for example with regard to material utilization and/or land occupation [197,222]. In line with Section 3.1, these additional environmental burdens mainly arise due to the oversizing of off-grid systems, which might be reduced with optimization and/or appropriate disposal of system components. Thus, there can be substantial environmental consequences when costs are the only metric considered within the analysis of off-grid energy systems. Therefore, we argue to consider additional metrics beyond costs and (operational) greenhouse gas emissions during the design phase of off-grid energy systems.

However, LCOEs are the only suitable metric to compare the economics of decentralized off-grid renewable energy systems at the moment, due to their coverage in most studies. The comparison with our LCOE overview in this review can, for example, prevent design errors in future studies, if authors find that their calculated LCOEs are too high or low. Therefore, our overview can be used to verify findings on off-grid systems, to assess where these systems might be deployed and how costs evolve.

**Acknowledgements**

This work was supported by the Helmholtz Association under the program "Energy System Design".

**CRediT-Statement**

Conceptualization, J.W.; Data curation: J.W. J.S.; Formal analysis: J.W.; Funding acquisition: D.S.; Investigation: J.W.; Methodology: J.W.; Software: J.S., P.K.; Supervision: D.S.; Validation: J.W.; Visualization: J.G., J.W.; Writing - original draft: J.W., M.H., T.T.; Writing - review & editing: R.M., M.H., J.G., T.T., J.S., P.K., L.K., J.L., D.S., J.W.